\documentclass[preprint,12pt,authoryear]{elsarticle}
\usepackage{amssymb}
\usepackage{times}
\usepackage{amsmath}
\usepackage{longtable}

\journal{New Astronomy}

\begin{document}

\begin{frontmatter}

\title{Timing of AB And eclipses}

\author[1]{V. S. Kozyreva}

\author[2]{M. A. Ibrahimov}

\author[3]{E. R. Gaynullina}

\author[3]{R. G. Karimov}

\author[3]{B. M. Hafizov}

\author[4]{B. L. Satovskii}

\author[5]{V. N. Krushevska}

\author[5]{Yu. G. Kuznyetsova}

\author[1]{A. I. Bogomazov}

\author[1]{T. R. Irsmambetova}

\author[2]{A. V. Tutukov}

\address[1]{M. V. Lomonosov Moscow State University, P. K. Sternberg Astronomical Institute, 13, Universitetskij prospect, Moscow, 119991, Russia}

\address[2]{Institute of astronomy, Russian Academy of Sciences, 48 Pyatnitskaya st., 119017, Moscow, Russia}

\address[3]{Ulugh Beg Astronomical Institute, Uzbek Academy of Sciences, 33, Astronomicheskaya ulitsa, Tashkent, 100052, Uzbekistan}

\address[4]{AstroTel Ltd., 1A, Nizhegorodskaya ulitsa, Moscow, 109147, Russia}

\address[5]{Main Astronomical Observatory, National Academy of Sciences of Ukraine, 27, Akademika Zabolotnoho ulitsa, Kyiv, 03680, Ukraine}

\begin{abstract}
This study aims timing the eclipses of the short period low mass binary star AB And. The times of minima are taken from the literature and from our observations in October 2013 (22 times of minima) and in August 2014 (23 times of minima). We find and discuss an inaccuracy in the determination of the types of minima in the previous investigation by \citet{li2014}. We study the secular evolution of the central binary's orbital period and the possibility of the existence of third and fourth companions in the system.
\end{abstract}


\end{frontmatter}

\section{Introduction}

AB And is a W UMa type eclipsing binary star found by \citet{guthnick1927}. The primary star's mass and radius are nearly solar values, for the secondary they are approximately half the solar values (e.g., \citealp{pych2004}, \\ \citealp{hasanzadeh2008}). Old photographic plates allowed to obtain times of minima of the AB~And light curve since 1902, such that now there is a set of observations longer than one hundred years. It consists of more than 1600 points according to the B.R.N.O. Eclipsing Binaries database\footnote{http://var2.astro.cz/ocgate/?lang=en}, including photographic, visual and photoelectric (CCD and photomultiplier) observations. 

The orbital period change in AB And was discovered by \citet{oosterhoff1950}, later the amount of observational data was growing, and different researchers gave different descriptions of the $(O-C)$ diagram as a superposition of some pulsations (e. g., \citealp{binnendijk1959,panchatsaram1981,demircan1994, kalimeris1994,borkovits2005}). In the latest paper \citet{li2014} found that there is a linear increase of the orbital period along with pulsations, and this increase was interpreted as mass transfer from the companion to the primary.

The evolution of low mass contact binaries was investigated by \citet{iben1984}. It was shown that the evolution of low mass ($\le 1.5 M_{\odot}$) binaries, components of which have convective envelopes, is driven mainly by loss of orbital angular momentum by the magnetic stellar wind. The initial separation of a main-sequence predecessor of W UMa has to be about
$6-10 R_{\odot}$. The final product of the evolution of W~UMa stars can be a blue straggler. Studies of the multiplicity of stellar systems also are very important for understanding the formation and evolution of stars (e. g., \citealp{tokovinin1999}).

\section{Observations of AB And and determinations of the times of minima}

AB And was observed using the 50 cm AMT-1 telescope with an Apogee Alta-U16M 4Kx4K CCD camera at the Maidanak observatory of the Ulugh Beg Astronomical Institute, Uzbek Academy of Sciences in October 2013 and in August 2014. We obtained several tens of thousands images in the Bessell {\it R} filter with exposure times ranging from 8 to 20 seconds, the typical value was 10 seconds. The source was continuously monitored for 5-7 hours per night. Bias and dark frames with appropriate exposures were made every night before and after the observations. Flat field frames were recorded for the twilight sky.

To process the data we used the aperture photometry method with the C-Munipack program\footnote{http://c-munipack.sourceforge.net}. The optimum aperture corresponded to the minimum of the standard deviation for differential magnitudes. We took 2Mass J15270686+3659270 and 2Mass J15272880+3647225 as reference stars. The aperture was constant during one night, its differences from night to night were insignificant. Maximum errors for a single exposure were in the range $0.0024^m-0.004^m$ for the different nights. Standard dark and flat field corrections were made. Original fits files were converted to text files that included the dependence of the AB And brightness with respect to a reference star on the heliocentric Julian date (HJD). Light curves of AB And were created on the basis of these data. In Figure 1 we show a sample light curve of AB And from our observations. In order to achieve the highest possible precision of the times of minima we used only the full light curves between their maxima.

We calculated 45 times of minima from our observational material, listed in Table 1. Using this high precision homogeneous material without inner systematic errors we attempted to find modulations with low amplitude in the same way as we did for CV Boo \citep{bogomazov2016}. The geometrical parameters of the system were taken from \citet{li2014}. $(O-C)$ values were calculated for the full set of times of minima (ours $+$ the B.R.N.O. database). We found that the $(O-C)$ values in our computations were different in comparison with the $(O-C)$ values in the paper by \citet{li2014} for the times of minima before 1930. 

There is a difficulty in the determination of the primary and secondary times of minima of W~UMa type light curves, because the system's orbit is circular, and the shapes of the minima are comparable in their duration and depth. If the system's relatively short orbital period \footnote{0.3318912 days for AB And according to General Catalogue of Variable Stars \citep{samus2017}.} is changing with time, times of minima can be calculated incorrectly. This is especially important for observations made in the distant past. The earliest times of minima were obtained in 1902-1910, then in 1926-1930, and since 1939 there is a continuous set of times of minima.

We computed the $(O-C)$ values for the times of minima and noticed that we obtained a maximum value for the $(O-C)$ value at the beginning of the observations (1902-1903) of $\approx 0.22$ days. In the paper by \citet{li2014} for this time this value is much greater and can reach up to 0.6 days. Since we used the same ephemerides as \citet{li2014} this discrepancy can only be due to a wrong determination of epoch of the minimum in comparison with the initial epoch. \citet{pribulla2001} obtained as ephemerides:

\begin{equation}
\text{Min}\ \text{I} = \text{HJD}2451534.25045+0.33189106\times E,
\end{equation}

\noindent where $\text{Min}\ \text{I}$ is the epoch of the primary minimum, $T_0=\text{HJD}2451534.25045$ is the initial epoch, $P_0=0.33189106$ days is the orbital period, $E$ is the number of orbital cycles since the initial epoch.

In the older paper of \citet{binnendijk1959} information about the kind of minimum (primary or secondary) was included.  The $(O-C)$ value calculated by \citet{li2014} can be obtained if they took one orbital period more before the initial epoch in 1902-1910, and in addition in 1926-1930 confused a primary with a secondary minimum. If the $(O-C)$ values are recalculated using the ephemerides by \citet{binnendijk1959}, our values of $E$ do not coincide with those of \citet{li2014}. For the first minimum HJD2416103.925 we took for the epoch number $E=-106753.5$, whereas \citet{li2014} took for this $E=-106754.5$.

\section{Periodic (O-C) variations}

We calculated the $(O-C)_1$ values by using all times of minima (ours $+$ the B.R.N.O. database) and Equation 1. The obtained $(O-C)_1$ values were fitted by the equation:

\begin{eqnarray}
\text{Min}\ \text{I}=(T_0+\Delta T_0)+(P_0+\Delta P_0)E+\frac{\beta}{2} E^2 \nonumber \\
+A_3\left(\sqrt{1-e_3^2}\sin E^* \cos \omega_3 + \cos E^* \sin \omega_3\right),
\end{eqnarray}

\noindent with the equality

$$
\frac{2\pi}{P_3}(t-T_3)=(E^*-e_3\sin E^*),
$$

\noindent here $\Delta T_0$ is the correction for the initial epoch, $\Delta P_0$ is the correction for the orbital period, $A_3=(a_3\sin i_3)/c$, $c$ is the speed of light, $a_3 \sin i_3$ is the projected semi-major axis of the binary's orbit around the center of masses of the assumed triple system, $e_3$ is its eccentricity, $\omega_3$ is its argument of periapsis, $E^*$ is its eccentric anomaly, $T_3$ is the time of the periastron passage by the third body, $t$ is the time. The numerical values of these parameters are listed in Table 2. A graphical representation of the result is shown in Figure 2 by the green curve.

The mass function is $f(M_3)=0.048 M_{\odot}$, so the lower limit of mass of the third companion is $M_3=0.4-0.5 M_{\odot}$. This value is much smaller than the lower mass limit obtained by \citet{li2014}, they estimated it as $\approx 2.5 M_{\odot}$.

The next step was to search for remaining smaller variations of $(O-C)$ relative to the green curve. The visual times of minima are not precise enough and can make only noise for this purpose. So, we used the photoelectric and photographic times of minima starting from the epoch HJD2429000. The earlier points calculated using photographic observations can have systematic errors (see Figure 2). We calculated values $(O-C)_2$ as the difference between observational times of minima and values calculated using Equation (2). The obtained quantities of $(O-C)_2$ were fitted by a light equation\footnote{See, e. g., \citet{kozyreva2005}. Formula (3) for the light equation, and Formula (5) for the mass function.} with the parameters shown in Table 3. A graphic presentation of the result is shown in Figure 3. These then imply the presence of a fourth companion. The mass function for this 4$^\text{th}$ body is $f(M_4)=0.00035 M_{\odot}$, and the lower limit of its mass is $M_4=0.1 M_{\odot}$.

To estimate the significance of our results we used a statistical method of \citet{stellingwerf1978} and calculated the value

\begin{equation}
\label{theta}
\theta=\frac{\sigma^2}{\sigma_0^2},
\end{equation}

\noindent where $\sigma_0$ and  $\sigma$ are the standard deviations. The smaller the $\theta$ the better is the agreement between the data and the theoretical fit.
 
For the ``Theory, combined'' curve in Figure 2: $\sigma_0=0.0304$ corresponds to the values of $(O~-~C)_1$, $\sigma=0.0037$ is corrected with the theoretical curve from Equation (2), and $\theta=0.015$. For the theoretical curve in Figure 3: $\sigma_0=0.00160$ corresponds to the values of $(O~-~C)_2$, $\sigma=0.00267$ is corrected with the light equation obtained using the parameters from Table 3, and $\theta=0.36$.

It is essential to note that an error in times of minima obtained using photographic observations can reach a value of up to $\approx 0.01$ days, therefore if the binary's orbital period is $\approx 0.3$ days the derived difference of $(O-C)\approx 0.2$ days between modern and old observations is impossible, so the secular change is real. We also tested possible variabilities for the data in Figure 3 with different periods and found that $\theta$ for them is $\geq 0.7$ (for the curve in Figure 3 this values is almost twice less). Nevertheless, future observations potentially are able to correct estimations of periods, especially after the next periastron passage.

\section{Discussion and conclusions}

We obtained 45 times of minima of AB And from our observations, and analysed them along with the times of minima from the B.R.N.O. database. Our work corrected claims made by \citet{li2014}, because we found an error in their determination of times of minima type for the old data. The results of the present study are similar in some respects with the results of \citet{hasanzadeh2008}: also in their paper there are two periodical modulations of the times of minima.

Using all available times of AB And minima we find that the central binary's orbital period shows a secular increase and two periodical modulations. Re-calculation of mass transfer rate using Equation (5) by \citet{li2014} with our value of $\beta$ from Table 1 gives $\dot M\approx 5\cdot 10^{-8}M_{\odot}$ per year. The mentioned modulations can be explained by the gravitational influence of a third and a fourth circumbinary bodies. The orbital periods and the lower limits of masses of these bodies are $P_3\approx 106$ years, $M_3\gtrsim 0.4 M_{\odot}$, $P_4\approx 59$ years, $M_4\gtrsim 0.1 M_{\odot}$ respectively.

The presence of two distant companions to the close binary AB And is not a great surprise.
The empirical initial distribution of components of binary stars over separations was described by Equation (22) of \citet{masevich1988}, page 110:

$$
dN= 0.2 d\log (a/R_{\odot}),\text{} 1\le \log (a/R_{\odot}) \le 6,
$$

\noindent To estimate the possible multiplicity we can treat this function as a probability to find a new companion in a multiple system. Selection effects could lead to missing of components of very close binaries or binaries with a faint companion, therefore binaries in reality can be triples/multiples. According to \citet{tokovinin2006} more than 96\% of binaries with orbital periods $\lesssim 3$ days have a tertiary companion. Our statements are in a good agreement with this radial velocity study.

Thus, the formation of several ``additional'' components to a close W~UMa type system even with a significant eccentricity of their orbits can be a natural product of the collapse of a rotating protostellar gas cloud. The orbits of the two new components of AB~And due to their large eccentricities (0.41 and 0.87) are very close to each other. The ratio of orbital periods is only about 1.8. The stability of such close orbits
deserves now a special attention, e. g. \citep{orlov2005,naoz2016}.

Also a magnetic mechanism \citep{applegate1992} can be an explanation for a modulation of the binary's orbital period. We re-calculated the values $\Delta L_1$ and $\Delta L_2$ (changes of the luminosity of components due the magnetic mechanism, we used Equations 6-10 by \citet{li2014} using our value of $A$ and obtained $\Delta L_1=0.42 L_{\odot}$ and $\Delta L_2=0.17 L_{\odot}$. The values of \citet{li2014} were inconsistent with the possible energy of the mechanism, whereas our values are significantly smaller. \citet{kalimeris1994} already discussed this issue and found arguments in favour as well as against Applegate's mechanism on a AB~And. Their luminosity variation in long time scales  was $\Delta (L_1+L_2)=0.18L_{\odot}$. \citet{hasanzadeh2008} attributed the second modulation of the orbital period to the magnetic mechanism. The possible causes of orbital period modulations can be (at least, in principle) clarified using future CCD photometry, radial velocity, astrometric and direct observations.

\clearpage

\begin{center}
\label{ourminima}
\begin{longtable}{@{}cc|cc|cc@{}}
\caption{AB And times of minima obtained in this study.}\\
\hline
HJD-2400000 & Min & HJD-2400000 & Min & HJD-2400000 & Min \\
\hline
56567.2122 & II & 56580.1555 & II & 56875.2067 & II \\
56567.3783 & I & 56580.3215 & I & 56875.3728 & I \\
56568.3737 & I & 56581.3168 & I & 56876.2025 & II \\
56569.2032 & II & 56589.2827 & I & 56876.3683 & I \\
56569.3693 & I & 56589.4486 & II & 56884.3336 & I \\
56571.1944 & II & 56590.2783 & I & 56886.3251 & I \\
56571.3608 & I & 56590.4443 & II & 56886.4908 & II \\
56572.1904 & II & 56871.2239 & II & 56887.3206 & I \\
56573.1859 & II & 56871.3902 & I & 56887.4868 & II \\
56573.3518 & I & 56872.2196 & II & 56888.3165 & I \\
56574.1817 & II & 56872.3858 & I & 56888.4825 & II \\
56574.3474 & I & 56873.2157 & II & 56889.3125 & I \\
56577.1688 & II & 56873.3818 & I & 56889.4786 & II \\
56577.3343 & I & 56874.2115 & II & 56890.3077 & I \\
56578.3297 & I & 56874.3776 & I & 56890.4740 & II \\
\hline
\end{longtable}
\end{center}

\clearpage

\begin{table}
\centering
\begin{minipage}{120mm}
\caption{Values of the parameters in Equation 2.}
\label{3b}
\begin{tabular}{@{}lll@{}}
\hline
Parameter & Value & Error \\
\hline
$\Delta T_0$, d & $0.041$ & $\pm 0.002$ \\
$\Delta P_0$, d & $1.1\cdot 10^{-7}$ & $\pm 0.7\cdot 10^{-7}$ \\
$\beta$, d/yr & $3.95\cdot 10^{-8}$ & $\pm 2.5\cdot 10^{-8}$ \\
$A_3$, d & $0.047$ & $\pm 0.002$ \\
$e_3$ & $0.41$ & $\pm 0.12$ \\
$P_3$, d & $38700$ & $\pm 1990$ \\
$\omega_3$, $^\circ$ & $290.0$ & $\pm 0.6$ \\
$T_3$, HJD & $2434950.6$ & $\pm 260.3$ \\
\hline
\end{tabular}
\end{minipage}
\end{table}

\begin{table}
\centering
\begin{minipage}{120mm}
\caption{Values of the parameters for the second periodical variation.}
\label{4b}
\begin{tabular}{@{}lll@{}}
\hline
Parameter & Value & Error \\
\hline
$A_4$, d & $0.0062$ & $\pm 0.0015$ \\
$e_4$ & $0.87$ & $\pm 0.2$ \\
$P_4$, d & $21650$ & $\pm 100$ \\
$\omega_4$, $^\circ$ & $19.5$ & $\pm 1.5$ \\
$T_4$, HJD & $2443047$ & $\pm 50$ \\
\hline
\end{tabular}
\end{minipage}
\end{table}

\clearpage

\begin{figure*}
\includegraphics[width=1.0\textwidth]{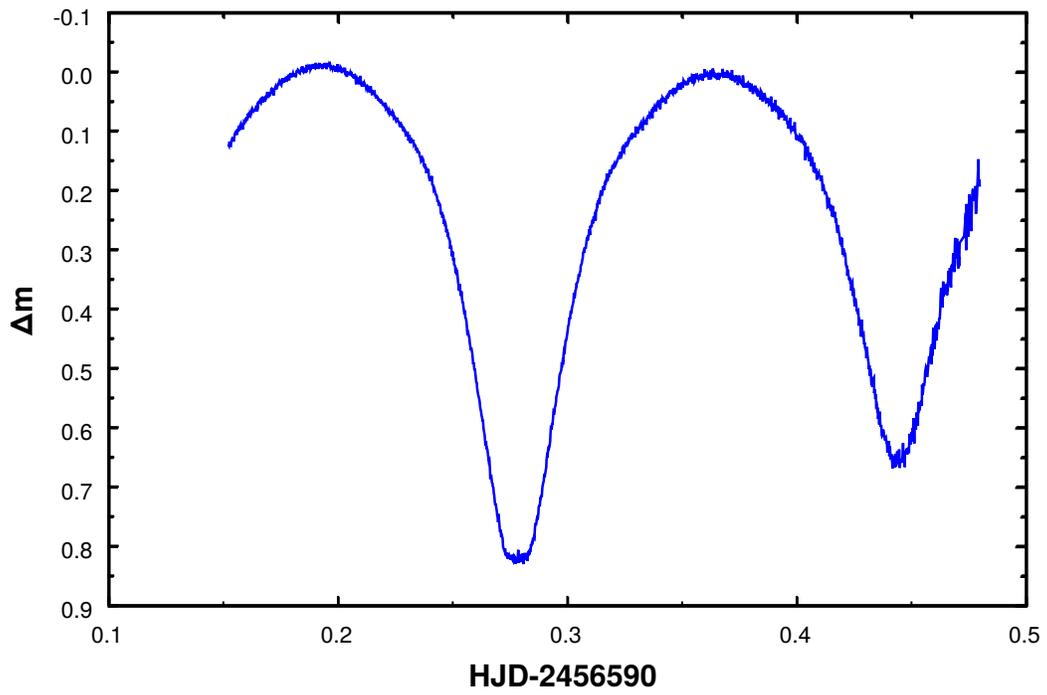}
\vspace{15pt} \caption{A sample light curve of AB And obtained in our observations in the Bessel {\it R} filter.}\label{f1}
\end{figure*}

\newpage

\begin{figure*}
\includegraphics[width=1.0\textwidth]{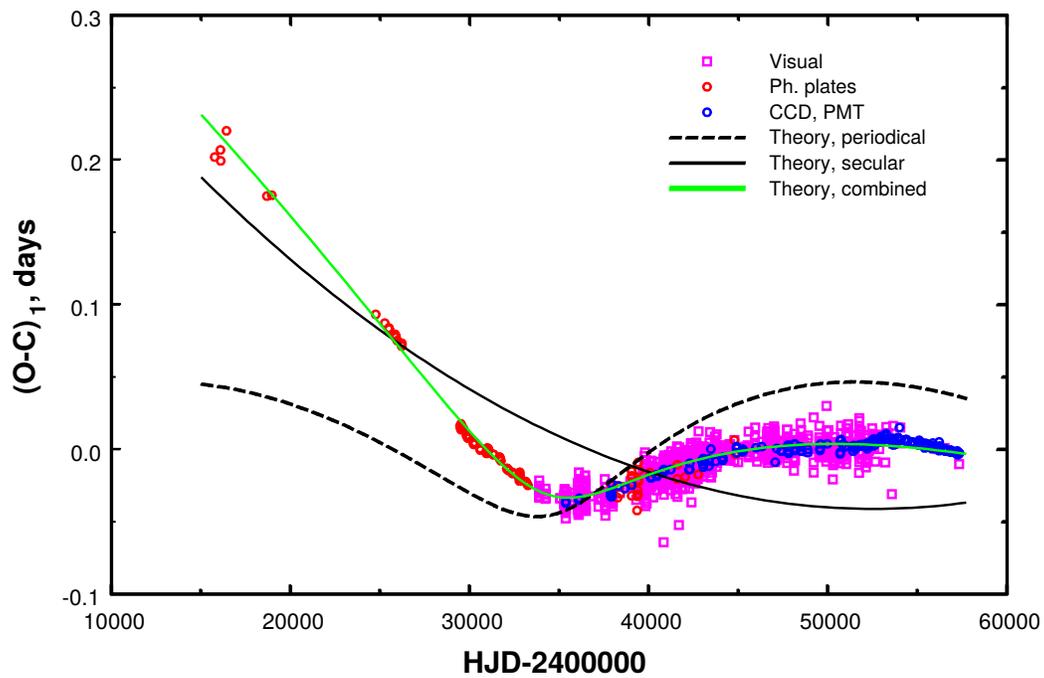}
\vspace{15pt} \caption{$(O-C)$ diagram, calculated using Equation (1). Curves represent a secular change of the system's orbital period, a periodical modulation of it, and the resulting combination of these two, represented by Equation (2). Circles depict photographic and photoelectric (CCD and PMT) observed times of minima, boxes depict visual data.}\label{f2}
\end{figure*}

\newpage

\begin{figure*}
\includegraphics[width=1.0\textwidth]{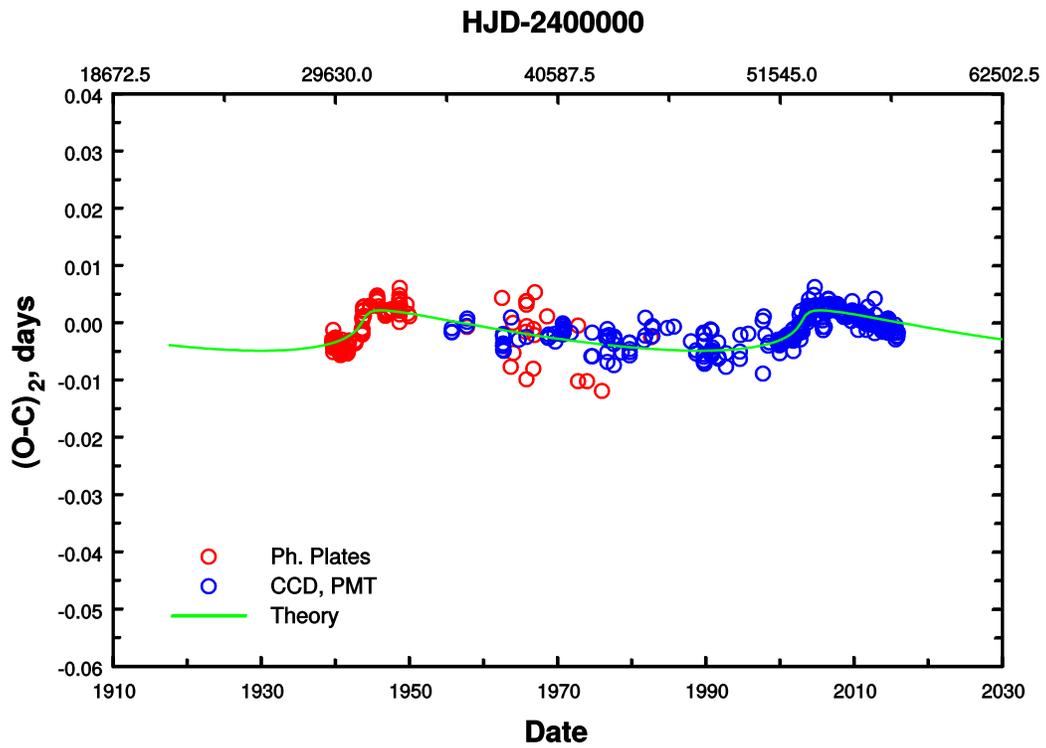}
\vspace{15pt} \caption{$(O-C)$ diagram for the smaller periodical variation after the subtraction of the combined theoretical modulation of Figure 2. The curve depicts the theoretical result; circles represent the photographic and photoelectric times of minima from HJD2429000.}\label{f3}
\end{figure*}

\end{document}